\documentclass[twocolumn,showpacs,preprintnumbers,amsmath,amssymb]{revtex4}
\usepackage{graphicx}
\usepackage{dcolumn}
\usepackage{bm}

\def\thalf{{\textstyle{\frac{1}{2}}}}

\def\tquar{{\textstyle{\frac{1}{4}}}}

\def\ttquar{{\textstyle{\frac{3}{4}}}}

\def\s{\sigma}

\def\pv{\vmg{\pi}}

\def\Bmunu{\vmg{B_{\mu\nu}}}
\def\b{\vmg{b}}
\newcommand{\vm}[1]{\mbox{\bf#1}}
\newcommand{\vmg}[1]{\mbox{\boldmath$#1$}}

\newcommand{\bq}    {\begin{equation}}
\newcommand{\eq}    {\end{equation}}
\newcommand{\bqr} {\begin{eqnarray}}
\newcommand{\eqr} {\end{eqnarray}}

\begin{document}

\title{A chiral lagrangian with broken scale:\\
a Fortran code with the thermal contributions of the dilaton field}

\author{Luca Bonanno}

\address{INFN Sezione di Ferrara, 44100 Ferrara, Italy}

\begin{abstract}
The Chiral Dilaton Model is a chiral lagrangian in which the breaking of scale invariance 
is regulated by the expectation value of a scalar field, called dilaton.
Here we provide a Fortran code \cite{epaps,ind}, as a tool to make calculations within 
the Chiral Dilaton Model at finite density and temperature. 
The calculations are improved respect to previous works~\cite{Bonanno:2007kh,Bonanno:2008tt}
by including the thermal contributions of the dilaton field.

\end{abstract}


\maketitle

\section{Introduction}
In the next years several Heavy Ion Collision (HICs) experiments at energies of the order of a 
few ten A GeV (as e.g. the ones proposed at facility FAIR at GSI \cite{Senger:2004jw}, at RHIC (Brookhaven) and
at the Nuclatron in Dubna) will probably start their activity. In these experiments the Equation of State 
(EOS) of matter will be tested at large density and/or temperature. 
It is therefore very important to provide, through theoretical investigations, a map of the 
``new'' regions which will likely be explored by the new experiments.

The Chiral Dilaton Model (CDM), developed by the group of the University of Minnesota
\cite{Heide:1993yz,Carter:1995zi,Carter:1996rf,Carter:1997fn} and largely
discussed in Refs.~\cite{Bonanno:2007kh,Bonanno:2008tt}, is a chiral hadronic model where 
chiral fields are present together with a dilaton field which
reproduces, at a mean field level, the breaking of scale invariance
taking place in QCD. The presence of two not-vanishing condensates, the chiral and
the dilaton condensates (the latter is connected to the gluon condensate), allows to study the restoration of both chiral
symmetry and scale invariance and the interplay between them.

In this work we provide a Fortran code \cite{epaps,ind} as a tool to explore the properties of the CDM.
In particular the code allows to compute the EOS, the masses and the mean values of the fields 
in a wide region of the density-temperature 
plane ($0.01\rho_0\leq \rho_B\leq 9.01\rho_0$ with $\rho_0=0.15 fm^{-3}$ and $0\leq T\leq 245$ MeV) and in a wide range 
of isospin asymmetries ($0.3\leq Z/A\leq 0.5$).
The code allows also to compute neutron star matter at T=0 (with electrons 
and muons in $\beta$-equilibrium) and T=0 pure neutron matter. A very small execution time is ensured
since the code makes a cubic spline interpolation of data previously computed.

There are two important improvements present in the code respect to the calculations provided in 
Refs.~\cite{Bonanno:2007kh,Bonanno:2008tt}.
The most important one is that the thermal fluctuations of the dilaton field are taken into account.  
This allows to study more in detail the regions of density and temperature where scale invariance is restored.
The second improvement concerns the technique used to compute the thermal averages, which is more reliable than the one used in 
Refs.~\cite{Bonanno:2007kh,Bonanno:2008tt} because it does not resort to any approximation.

The paper is structured as follows: in Sec II the lagrangian of the CDM is described, in Sec III the technique used to compute the
thermal averages is presented, in Sec. IV the procedure used to make calculations 
of the model is described. Finally in Sec. V we conclude the paper by showing some applications of the code.

\section{The Chiral Dilaton Model}
The lagrangian of the CDM reads:

\begin{eqnarray}
{\cal L}&=&\thalf\partial_{\mu}\sigma\partial^{\mu}
\sigma+\thalf\partial_{\mu}\vmg{\pi}\cdot\partial^{\mu}\vmg{\pi}
+\thalf\partial_{\mu}\phi\partial^{\mu}\phi
-\tquar\omega_{\mu\nu}\omega^{\mu\nu}\nonumber\\
&-&\tquar\Bmunu\cdot\Bmunu
+\thalf G_{\omega\phi}\phi^2 \omega_\mu\omega^\mu 
+\thalf G_{b\phi}\phi^2 \b_\mu\cdot\b^\mu\nonumber\\ 
&+&[(G_4)^2\omega_\mu\omega^\mu]^2-{\cal V}\\
&+&\bar{N}\left[\gamma^\mu(i\partial_{\mu}-g_\omega\omega_\mu
-\thalf g_{\rho}\b_{\mu}\cdot\vmg{\tau})
-g\sqrt{\s^2+\pv^2} \right] N \label{lb}\nonumber
\end{eqnarray}
where the potential is:
\begin{eqnarray}
{\cal V}&=& B\phi^4
\left(\ln\frac{\phi}{\phi_0}-\frac{1}{4}\right)
\hspace{-.73mm}-\hspace{-.73mm}\thalf B\delta\phi^4
\ln\frac{\sigma^2+\vmg{\pi}^2}{\sigma_0^2}\nonumber\\
&+&\hspace{-.74mm}\thalf B\delta \zeta^2\phi^2\!\!\left[\sigma^2
+\vmg{\pi}^2-\frac{\phi^2}{2\zeta^2}\right]-\ttquar\epsilon_1'\label{lm}\\
&-&\tquar\epsilon_1'\left(\frac{\phi}{\phi_0}\right)^{\!2}
\left[\frac{4\sigma}{\sigma_0}-2\left(\frac{\sigma^2
+\vmg{\pi}^2}{\sigma_0^2}\right)-\left(\frac{\phi}{\phi_0}\right)^{\!2}
\,\right]\,.  \label{pot}\nonumber
\end{eqnarray}
Here $\sigma$ and $\vmg{\pi}$ are the chiral fields, $\phi$ the
dilaton field, $\omega_{\mu}$ the vector meson field and ${\bf
b}_{\mu}$ the vector-isovector meson field, introduced in order to
study asymmetric nuclear matter. The field strength tensors are
defined by $F_{\mu\nu}=\partial_{\mu}\omega_{\nu}-\partial_{\nu}\omega_{\mu}$,
$\vm{B}_{\mu\nu}=\partial_{\mu}\vm{b}_{\nu}-\partial_{\nu}\vm{b}_{\mu}$.
In the vacuum ${\phi}={\phi_0}$, ${\sigma}={\sigma_0}$ and
${\vmg{\pi}}=0$.  The $\omega$ and $\rho$ vacuum masses are
generated by their couplings with the dilaton field so that
$m_{\omega}=G_{\omega\phi}^{1/2} \phi_0$ and
$m_{\rho}=G_{\rho\phi}^{1/2} \phi_0$. Moreover
$\zeta={\phi_0}/{\sigma_0}$, B and $\delta$ are constants and
$\epsilon_1'$ is a term that breaks explicitly the chiral invariance
of the lagrangian. 
Finally, the values of the parameters used in calculations (listed in table~1 of Ref.~\cite{Bonanno:2008tt}) 
were determined in Ref.~\cite{Carter:1995zi} by fitting 
the properties of nuclear matter and finite nuclei.

\section{Thermal averages}

In finite temperature mean field approximation, when the temperature is large enough to become comparable 
to the masses of the meson fields it is mandatory to take into account the
thermal fluctuations of those fields. In Ref.~\cite{Bonanno:2008tt} we 
considered the thermal fluctuations of the chiral fields and of the 
vector mesons, but the thermal fluctuation of the dilaton field was not considered due to the 
large value of the glueball mass (about 1.6 GeV in the vacuum). 
In the present work we extended the previous calculations by including also the fluctuation of the dilaton field.

As already discussed in Ref.~\cite{Bonanno:2008tt}, it is not trivial
to compute the thermal averages of a quantity depending on the fluctuating fields, in particular when
those fields do not appear in polynomial form. 
Due to the complicate structure of the potential (\ref{pot}), almost all quantities depending on the scalar fields 
are difficult to average (notice that the terms containing only the vector fields are not problematic).

In Ref.~\cite{Bonanno:2008tt} we presented two different techniques to compute the thermal averages: 
the technique developed by the authors of the CDM \cite{Carter:1996rf,Carter:1997fn}, which we mainly 
adopted in Refs.~\cite{Bonanno:2007kh,Bonanno:2008tt}, and a technique developed in a more recent paper by Mocsy and Mishustin 
\cite{Mocsy:2004ab}.

The first technique \cite{Carter:1996rf,Carter:1997fn} was developed to treat in a compact way the chiral invariant term 
$(\sigma^2+\pv^2)$, which appears frequently into the lagrangian. To this purpose two approximations were introduced:
the fluctuations of the chiral fields are assumed to be equal and the term $\bar\sigma\Delta\sigma$, supposed to be 
small, is expanded up to the fourth order.
The second technique \cite{Mocsy:2004ab}, instead, does not resort 
to any approximation and the thermal fluctuations 
are taken to all orders (notice that this request is not satisfied in the previous technique, because 
the expansion of the term $\bar\sigma\Delta\sigma$ is truncated to the fourth order). 
Moreover the technique of Ref.~\cite{Mocsy:2004ab} is a general procedure extensible 
to a generic number of fields, where each thermal fluctuation is treated as a independent quantity (this allows to
treat separately each isospin component of the pion, enabling, in this way, to compute
the isospin splitting of the pion mass).

For these reasons and since we are interested to include in the calculation the thermal fluctuation of the dilaton field we adopt in
the present calculation the technique by Mocsy and Mishustin \cite{Mocsy:2004ab}.
Extending the averaging procedure of Ref.~\cite{Mocsy:2004ab} by including the fluctuation of the dilaton field 
and treating the isospin components of the pion as independent of each other, we obtain that the thermal 
average of a function $A(\phi,\sigma,\pi_0,\pi_+,\pi_-)$ reads:
\bqr
&\langle A(\bar\phi+\Delta\phi,\bar\sigma+\Delta\sigma,\pi_0,\pi_+,\pi_-) \rangle = \nonumber\\
&\int\limits_{-\infty}^\infty\,dz_{\phi} P(z_{\phi},\langle\Delta\phi^2\rangle)
\int\limits_{-\infty}^\infty\,dz_{\sigma} P(z_{\sigma},\langle\Delta\sigma^2\rangle)\nonumber\\
&\times\int\limits_{-\infty}^\infty\,dz_{\pi_0} P(z_{\pi_0},\langle\pi_0^2\rangle) 
\int\limits_{-\infty}^\infty\,dz_{\pi_+} P(z_{\pi_+},\langle\pi_+^2\rangle)\\ 
&\times\int\limits_{-\infty}^\infty\,dz_{\pi_-} P(z_{\pi_-},\langle\pi_-^2\rangle) 
A(\bar\phi+z_{\phi},\bar\sigma+z_{\sigma},z_{\pi_0},z_{\pi_+},z_{\pi_-})\nonumber
\label{step3}
\eqr
where 
\bqr
P(z_i,\langle\Delta_i^2\rangle)=(2\pi \langle\Delta_i^2\rangle)^{-1/2} \exp\left(-\frac{z_i^2}{2 \langle\Delta_i^2\rangle}\right)
\label{weights}
\eqr
is a gaussian weighting function depending on the thermal fluctuation $\langle\Delta_i^2\rangle$ of the i-th scalar field.
Notice that this averaging procedure ensures the equivalence between the
resolution of the field equations and the minimization of the thermodynamical potential (this equivalence was only approximate with
the procedure developed by the authors of the CDM), as it has been proved in Ref.~\cite{Mocsy:2004ab}.

\section{Finite temperature calculations}
In this section we summarize the procedure adopted to make calculations at finite temperature.

The symmetries of the lagrangian lead to the conservation of the baryon number and of the isospin. 
Depending on the thermodynamical ensemble used, the conservation of these charges can be expressed by fixing their chemical potentials 
or their densities. Since we use the canonical ensemble we fix the baryon density $\rho_B$, 
the total charge per baryon $Z/A$ (that is equivalent to fix the total isospin density) and the temperature $T$.

Differently from what done in Ref.~\cite{Bonanno:2008tt} (where the mean values of $\sigma$ and $\phi$
were evaluated by minimizing the thermodynamical potential), here the calculation consists entirely on
the numerical resolution of a system of equations in the following variables:\\
- the nucleon chemical potentials $\mu_p$ and $\mu_n$. \\
- the thermal fluctuations $\langle\Delta\phi^2\rangle$, $\langle\Delta\sigma^2\rangle$
, $\langle\pi_0^2\rangle$, $\langle\pi_+^2\rangle$, $\langle\pi_-^2\rangle$, 
$\langle\Delta\omega_{\mu}\Delta\omega^{\mu}\rangle$, $\langle\Delta b_{0\mu}\Delta b^{\mu}_0\rangle$, 
$\langle\Delta b_{+\mu}\Delta b^{\mu}_+\rangle$, $\langle\Delta b_{-\mu}\Delta b^{\mu}_-\rangle$\\
- the mean field values $\bar\phi$, $\bar\sigma$, $\omega_0$, $b_0$.\\
As it will be shown in the following subsections, 2 of the 15 equations come from the conservation of the baryon and isospin charges 
, 9 are the self-consistency relations for the thermal fluctuations and the remaining 4 are the field equations.

\subsection{Conserved charges} 

In order to fix the total baryon charge and the total electric charge one has to solve the following 2 equations:

\bqr
\rho_B&=&\rho_p(m_N^*(\bar\phi_i,\langle\Delta\phi_i^2\rangle),\mu_p^*,T)\nonumber\\
&+&\rho_n(m_N^*(\bar\phi_i,\langle\Delta\phi_i^2\rangle),\mu_n^*,T)\nonumber\\
\rho_B\frac{Z}{A}&=&\rho_p(m_N^*(\bar\phi_i,\langle\Delta\phi_i^2\rangle),\mu_p^*,T)\nonumber\\
&+&\rho_{\pi_+}(m_{\pi_+}^*(\bar\phi_i,\langle\Delta\phi_i^2\rangle),\mu_{\pi_+}^*,T)\\
&-&\rho_{\pi_-}(m_{\pi_-}^*(\bar\phi_i,\langle\Delta\phi_i^2\rangle),-\mu_{\pi_+}^*,T)\nonumber\\
&+&\rho_{\rho_+}(m_{\rho}^*(\bar\phi_i,\langle\Delta\phi_i^2\rangle),\mu_{\rho_+}^*,T)\nonumber\\
&-&\rho_{\rho_-}(m_{\rho}^*(\bar\phi_i,\langle\Delta\phi_i^2\rangle),-\mu_{\rho_+}^*,T)\nonumber\label{fixed}
\eqr

where the index i runs on the scalar and vector fields, $\rho_i(m,\mu,T)$ are density integrals,
and the effective chemical potentials of the nucleons which enter the thermodynamical integrals 
are related to the standard ones as follows:
\begin{eqnarray}
\mu_p^*\,&=&\, \mu_p-g_{\omega}\omega_0-\thalf
g_{\rho}b_0\nonumber\\
\mu_n^*&\,=&\, \mu_n-g_{\omega}\omega_0+\thalf
g_{\rho}b_0\;. \label{potchim}
\eqr
The effective masses of the mesons are given by:
\bqr
m_i^{*2}=\pm\left\langle \frac{\partial^2 \cal L}{\partial \Delta_i^2}\right\rangle
\label{masses}
\eqr
where the minus sign stays for the scalar mesons and the plus sign stays for the vector mesons. Finally the nucleon mass reads:
\bqr
m_N^*=g\left\langle\sqrt{\sigma^2+\pv^2}\right\rangle
\label{mnu}
\eqr

Notice that the effective masses depend on the mean values and on the fluctuations of the fields which are needed 
to compute the thermal averages (\ref{masses}) and (\ref{mnu}).

The chemical potentials associated to the charged iso-vector mesons
are related to the chemical potentials of the nucleons by the relation \cite{Bonanno:2008tt}: 
\begin{eqnarray}
\mu_{\pi^+} = \mu_{\rho^+}  = \mu_{p}-\mu_{n}  
\end{eqnarray}
and the effective chemical potentials, which enter the thermodynamical integrals are
given by:
\bq
\mu_{\pi^+}^*= \mu_{\rho^+}^*=\mu_{p}^*-\mu_{n}^*\, .\label{potchimmes}
\eq

\subsection{Self-consistency relations for the thermal fluctuations}

The thermal fluctuation for a generic meson field is given by:

\bqr
\langle\Delta_i^2\rangle=\frac{n_i}{2\pi^2}\int\limits_{0}^\infty\,dk\frac{k^2}{\sqrt{k^2+m_i^{*2}}}
\frac{1}{e^{\beta(\sqrt{k^2+m_i^{*2}}-\mu_i^*)}-1}\label{fluc}
\eqr

where $m_i$ is its effective mass, $\mu_i^*$ its effective chemical potential and 
$n_i$ the degeneracy factor:
\bqr n_i=
\left\{\begin{array}{ll}
1 & \textrm{for the $\phi$, $\sigma$, $\pi_0$, $\pi_+$ and $\pi_-$ mesons}\\
-3 & \textrm{for the $\omega_{\mu}$, $b_{0\mu}$, $b_{+\mu}$ and $b_{-\mu}$  mesons}\\
\end{array}\right.
\label{dfactor}
\eqr

In order to compute the thermal fluctuation of a meson field (\ref{fluc}) one needs the value of its
effective mass, which in turn depends on the thermal fluctuations and on the mean values
of all the fields entering the expression (\ref{masses}).
This leads to a system of 9 self-consistent equations which reads: 

\bqr
\left\{\begin{array}{ll}
\ldots \\
\ldots \\
\langle\Delta_i^2\rangle=\frac{n_i}{2\pi^2}\int\limits_{0}^\infty\,dk\frac{k^2}{e_{i}^{*}\left(\bar\phi_j,\langle\Delta_j^2\rangle\right)}
\frac{1}{\exp\left[\beta\left(e_i^{*}\left(\bar\phi_j,\langle\Delta_j^2\rangle\right)-\mu_i^*\right)\right]-1}\\
\ldots \\
\ldots
\end{array}\right.
\label{syst}
\eqr
where the indexes i and j run on the scalar fields $\phi$, $\sigma$, $\pi_0$, $\pi_+$, $\pi_-$ and on the vector fields 
$\omega_{\mu}$, $b_{0\mu}$, $b_{+\mu}$, $b_{-\mu}$.
In the case of isospin symmetric nuclear matter, the thermal fluctuations of the isospin components 
of the pion and of the $\rho-$~meson are equal, so the system~(\ref{syst}) reduces to a system of 5 equations.

\subsection{Field equations} 

The mean field values $\bar\phi$, $\bar\sigma$, $\omega_0$ and $b_0$ 
are computed by solving their field equations, which are:
\bqr
0&=&\bigg\langle 4 B_0 \chi^3\ln\chi-
2B_0\delta\chi^3\ln\left(\frac{\sigma^2+\pv^2}{\sigma_0^2}\right)\nonumber\\
&-&B_0\delta\chi\frac{\sigma^2+\pv^2}{\sigma_0^2}-b_0\delta\chi^3
+\epsilon_1'\chi\left(\frac{\sigma^2+\pv^2}{\sigma_0}\right)
+\epsilon_1'\chi^3\bigg\rangle\nonumber\\
&-&2\epsilon_1'\bar\chi\frac{\bar\sigma}{\sigma_0}
-m_{\omega}^2\bar\chi(\omega_0^2+\langle\Delta\omega_{\mu}\Delta\omega^{\mu}\rangle)\nonumber\\
&-&m_{\omega}^2\bar\chi(b_0^2+\langle\Delta\b_{\mu}\cdot\Delta\b^{\mu})\rangle \;,\\
0&=&\bigg\langle \frac{g \sigma_0^2\sigma}{\sqrt{\sigma^2+\pv^2}}\bigg\rangle 
\big[\rho_{Sp}(m_N^*(\bar\phi_i,\langle\Delta\phi_i^2\rangle),\mu_p^*,T)\nonumber\\
&+&\rho_{Sn}(m_N^*(\bar\phi_i,\langle\Delta\phi_i^2\rangle),\mu_p^*,T)\big]-
\bigg\langle B_0\sigma_0^2\delta\chi^4\left(\frac{\sigma}{\sigma^2+\pv^2}\right)\nonumber\\
&+&(B_0\delta+\epsilon_1')\chi^2\sigma
-\epsilon_1'\chi^2\sigma_0\bigg\rangle\nonumber\\
0&=&4 G_4^4\omega_0^3+m_{\omega}^2(\bar\chi^2+\langle\Delta\chi^2\rangle)\omega_0-g_{\omega}\rho_B\nonumber\\
0&=&m_{\rho}^2(\bar\chi^2+\langle\Delta\chi^2\rangle) b_0-g_{\rho}\rho_3\nonumber\label{eqmot}
\end{eqnarray}

where $\rho_{Sp}$ and $\rho_{Sn}$ are the scalar densities of protons and neutrons, respectively, $\rho_3=(\rho_p-\rho_n)/2$ and
$\chi=\phi/\phi_0$.

\begin{figure}[t!]
\includegraphics[scale=0.5]{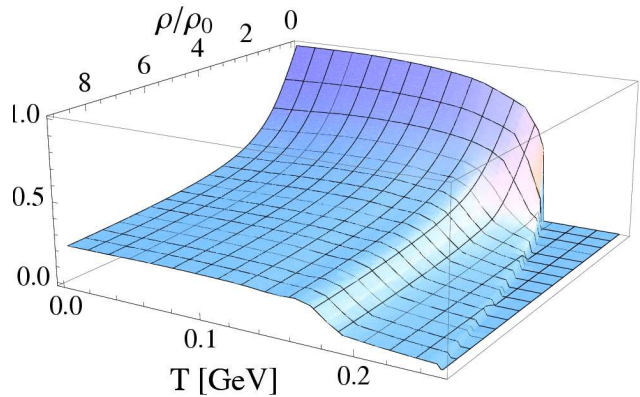}
\caption{\label{ni}
\footnotesize 3D-plot of the ratio $\bar\sigma/\sigma_0$ as a function of baryon density and temperature. Here Z/A=0.5}
\end{figure}
\begin{figure}[t!]
\includegraphics[scale=0.5]{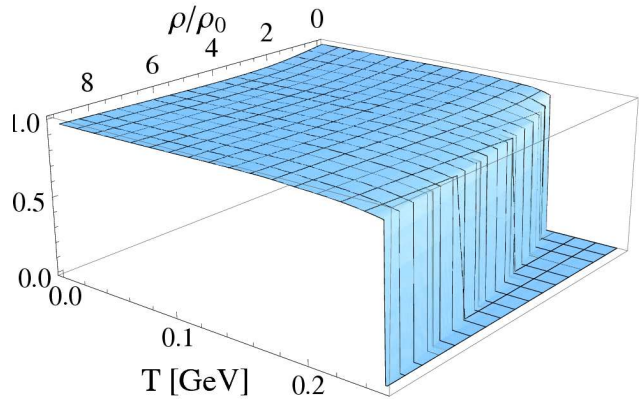}
\caption{\label{chi}
\footnotesize 3D-plot of the ratio $\bar\phi/\phi_0$ as a function of baryon density and temperature. Here Z/A=0.5}
\end{figure}

\section{Applications} 

The CDM code \cite{epaps,ind} can be used for many applications since it allows to compute many 
useful quantities in a wide range of temperatures,
baryon densities and isospin asymmetries. One of the most interesting applications is the study of chiral and scale phase transitions. 
To this purpose we used the code to build a phase diagram, as an extension of 
Fig.~8 of Ref.~\cite{Bonanno:2008tt}, by computing $\bar\sigma/\sigma_0$ (see Fig.~\ref{ni}) and 
$\bar\phi/\phi_0$ (see Fig.~\ref{chi}) as functions of T and $\rho_B$ in the case of isospin symmetric matter.
Notice that at large temperatures $\bar\sigma/\sigma_0$ sharply decreases reaching a plateau where it assumes very small values. 
This plateau is a region where chiral symmetry is strongly restored, while scale invariance is still broken. 
When temperature becomes even larger scale invariance restores with a first 
order transition ($\bar\phi/\phi_0$ vanishes) \footnote{We are working in a mean field approximation and it is known that 
the order of a phase transition cannot be determined in that approximation.}, dragging together the restoration 
of chiral symmetry ($\bar\sigma/\sigma_0$ vanishes). The code allows also to study the role of the isospin in chiral and
scale phase transitions.

It is necessary to point out that the results obtained here do not appear very different 
by those presented in Refs.~\cite{Bonanno:2007kh,Bonanno:2008tt} (although some differences are observed in the 
isospin splitting of the pion mass). The thermal fluctuation of the dilaton 
field, indeed, does not play a fundamental role. This is due to the value of the dilaton 
mass which remains large (about 1 GeV) also in the density-temperature region where the scale invariance restores. 

In the future it would be interesting to include the hyperons in calculations, through an extension of the model in SU(3).
This direction is actually explored by the Frankfurt group~\cite{Dexheimer:2008ax}. The most interesting extension is the inclusion
of the quark degrees of freedom, what will be done in the next works.

\section{Acknowledgments}
It is a pleasure to thank Alessandro Drago for his useful advices. 
     
\section*{References}
\bibliography{biblio.bib}

\end{document}